\documentclass[reprint,aps,amsmath,amssymb,twoside,prd,showkeys,superscriptaddress]{revtex4-1}
\pdfoutput=1
\usepackage{verbatim}
\usepackage{comment}
\usepackage{graphicx}
\usepackage[T1]{fontenc}
\usepackage{graphicx}
\usepackage{epsfig}
\usepackage{bm}
\usepackage{tensor}
\usepackage{amsfonts}
\usepackage{lipsum, babel}
\usepackage{amssymb}
\usepackage{float}
\usepackage{amsmath}
\usepackage{subfigure}
\usepackage{dcolumn}
\usepackage{cancel}
\usepackage[colorlinks]{hyperref}
\usepackage[utf8]{inputenc}

\usepackage[usenames,dvipsnames]{color}
\hypersetup{
     breaklinks=true,
    pdfstartview={FitH},    
    colorlinks=true,       
    linkcolor=blue,          
    citecolor=red,        
    filecolor=magenta,      
    urlcolor=blue,           
    anchorcolor=green,      
    linktocpage=true
}

\def\doi{http://doi.org}

\newcommand{\be}{\begin{equation}}
\newcommand{\ee}{\end{equation}}

\newcommand{\HCd}{\mathcal{H}}

\def\HCdt0{\tilde{\HCd}_{0}}

\newcommand{\affcam}{DAMTP, Centre for Mathematical Sciences, University of Cambridge, Wilberforce Road, Cambridge CB3 0WA, United Kingdom}
\newcommand{\affcamast}{Kavli Institute of Cosmology (KICC), University of Cambridge, Madingley Road, Cambridge, CB3 0HA, United Kingdom}
\newcommand{\affmColCam}{Queens’ College, Cambridge, CB3 9ET, United Kingdom}

\begin{document}

\title{Testing modified gravity via Yukawa potential in two body problem: Analytical solution and observational constraints}
\author{David Benisty}
\email{db888@cam.ac.uk}
\affiliation{\affcam}\affiliation{\affcamast}\affiliation{\affmColCam}
\begin{abstract}
Many alternative theories of gravity screens a Yukawa-type potential. This article shows Keplerian-type parametrization as a solution of Yukawa type potential accurate equations of motion for two non-spinning compact objects moving in an eccentric orbit. A bound from the solar system is presented.
\end{abstract}
\maketitle
\section{Introduction}
\label{sec:Introduction}
Cosmological measurements from the last few decades shows that General theory of Relativity (GR) is not the complete solution for gravity theories. The measurements from the Type Ia supernova \cite{Pan-STARRS1:2017jku}, Baryon Acoustic Oscillations (BAO) \cite{Addison:2013haa,Aubourg:2014yra,Cuesta:2014asa,Cuceu:2019for} and the Cosmic Microwave Background (CMB) \cite{Planck:2018vyg} give a strong evidence at least for one modification beyond GR, which is the Cosmological Constant $\Lambda$ \cite{Perlmutter:1998np,Weinberg:1988cp,Lombriser:2019jia,Copeland:2006wr,Frieman:2008sn,Riess:2019cxk}. However, the question whenever $GR\,+\,\Lambda$ is the final theory of gravity or a small part from a bigger theory remains an open question \cite{Capozziello:2011et}.




In order to constraint other theories of gravity there are many tests from the laboratory \cite{Vazza:2017qge} to compact objects \cite{2016AA...594A.107K,Banik:2018ydl}. The two-body solution for alternative theories also yields a strong constraint from solar system, the galactic star center \cite{Yu:2016nzn,2016ApJ83017B,Abuter:2018drb,2009ApJ707L114G,Do:2019txf,Abuter:2020dou,Amorim:2019hwp,2017ApJ...845...22P,Will:2018ont,Will:1997bb,Scharre:2001hn,Moffat:2005si,Zhao:2005zq,Bailey:2006fd,Deng:2009tg,Barausse:2012da,Borka:2012tj,Enqvist:2013tsa,Borka:2013dba,Berti:2015itd,Borka:2015vqa,Zakharov:2016lzv,Zhang:2017srh,Dirkes:2017ecu,Pittordis:2017byg,Hou:2017cjy,Nakamura:2018yaw,Banik:2018ydl,Dialektopoulos:2018iph,Kalita:2018ubo,Will:2018ont,Banik:2019zme,Pittordis:2019kxq,Nunes:2019bjq,Anderson:2019eay,Gainutdinov:2020bbv,Bahamonde:2020bbc,Banerjee:2020rrd,Ruggiero:2020yoq,Okcu:2021oke,deMartino:2021daj,DellaMonica:2021xcf,DAddio:2021xsu} and others systems \cite{Henrichs:2020cae}.

One of the simplest way to test gravity is to constrain the
existence of a Yukawa additional term, such that the one particle Newtonian potential assumes the form:
\begin{equation}
V(r) = \frac{G M}{r}\left(1 + \alpha \, e^{-m \cdot r}\right),
\end{equation}
where $M$ is the total mass of the system, $G $ is the Newtonian constant and $r$ is the separation between the two objects. For the Yukawa couplings: $\alpha$ is the Yukawa strength and $m$ is the Yukawa mass. For $\alpha$ goes to zero the Yukawa interaction is reduced into the Newtonian one. For $m$ goes to zero the Newtonian constant is modified to $G \left(1 + \alpha\right)$. Such a correction arises whenever the force is mediated by a scalar particle of mass $m$. Therefore testing gravity is equivalent to testing the existence of a fifth force of scalar nature.

Analysis of Yukawa potential in two body problem is studies in different approaches \cite{Pricopi:2016yfe,Edwards:2017ndv,2018InJPh..92..197M,Cavan:2020xnw,Martz:2019zba,Berge:2018htm,DeMartino:2018yqf,DeLaurentis:2018ahr}. In this {\it{letter}} we derive an analytical solution for the Yukawa potential using the equations of motion for compact binaries moving in eccentric orbits. We employ similar techniques which allowed to obtain a simple parametrization for the solution of Post Newtonian accurate equations of motion for compact binaries in eccentric orbits \cite{Memmesheimer:2004cv}. The approach yields to a close elegant solution and we constraint these parameters with different astrophysical systems.

The plan of work is as follows: Section \ref{sec:anaSol} introduces an analytical solution for the Yukawa potential in the Newtonian limit. Section \ref{sec:relCor} solves the relativistic correction to the Yukawa potential. Section \ref{sec:dataCom} performs data comparison with the solar system. Finally, section \ref{sec:dis} summarizes the results.

\begin{figure}[t!]
 	\centering
\includegraphics[width=0.45\textwidth]{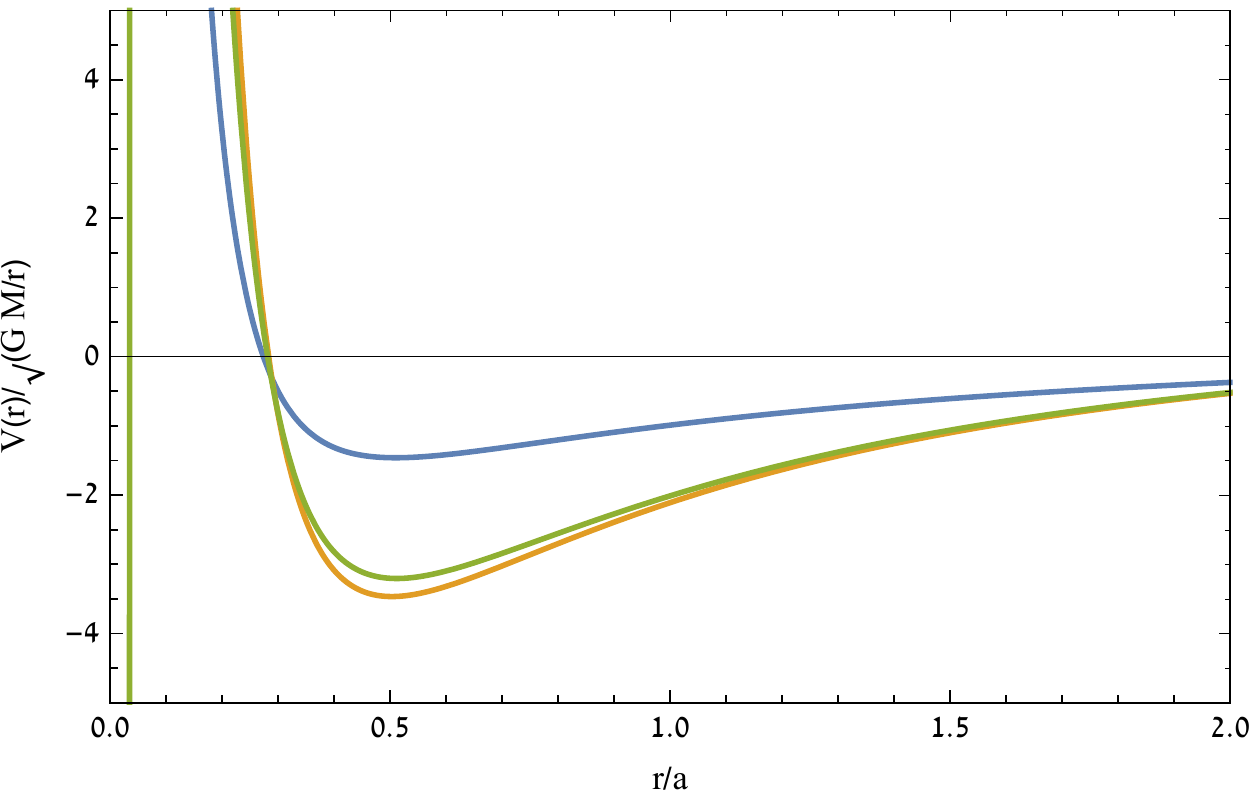}
\caption{\it{An effective comparison between the Newtonian potential (blue), the low energy Yukawa potential (orange) and Yukawa potential with a relativistic correction (green). }}	
\label{fig:potenYuk}
	\end{figure}

\section{Mean and true Keplerian motion}
\label{sec:anaSol}
We begin by summarizing some known solutions about the two body problem. The Keplerian parametrization for Newtonian motion is analytically solved in celestial mechanics \cite{brouwer1961methods}. The conservation of the energy and the angular momentum reads:

\begin{equation}
\epsilon = \frac{1}{2}v^2 -\frac{G M }{r}, \quad l = r^2 \dot{\theta},
\label{eq:NewEqu}
\end{equation}

\begin{figure}[t!]
 	\centering
\includegraphics[width=0.38\textwidth]{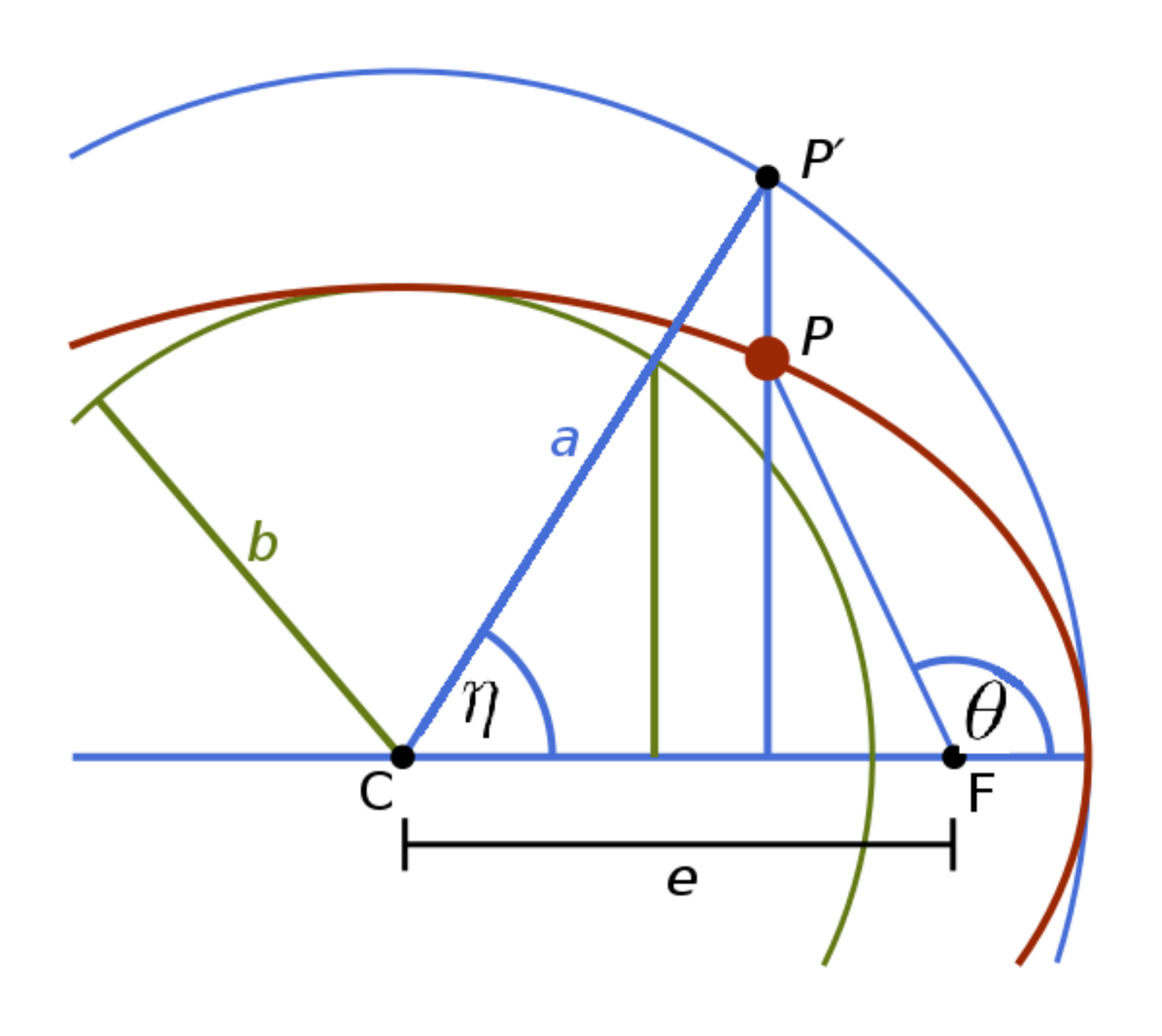}
 	\label{fig:trueAnomMeanAnom}
 \caption{\it{ {A demonstration for the difference between mean anomaly $\eta$ and the true anomaly $\theta$. The red curve is the true motion of the celestial body and the blue curve is the theoretical circle that surrounds the elliptical motion. The picture were taken from the link: \footnote{https://www.pngwing.com/en/free-png-shacp}.}}}
\end{figure}

where $\epsilon$ is the total energy per the reduced mass, $l$ is the total angular momentum per the reduced mass, $r$ is the separation and the dot sing is the derivative with respect to time. The total velocity in polar coordinates is:
$v^2 = \dot{r}^2 + \dot{\theta}^2r^2 = \dot{r}^2 + \frac{l^2}{r^3}$. In order to solve the problem it is useful to parametrize the separation as:
\begin{equation}
r/a = 1-e_r \cos\eta,
\label{eq:conSep}
\end{equation}
the exact solution for the Eq.~(\ref{eq:NewEqu}) is described by the well-known relations:
\begin{subequations}
\begin{equation}
\frac{2 \pi}{T} \left(t - t_0 \right) = \eta - e \sin \eta,
\label{eq:meanAnomNew}
\end{equation}
\begin{equation}
\theta = \tilde{\nu} \equiv 2\tan^{-1}\left[\sqrt{\frac{1+e}{1-e}} \tan \frac{\eta}{2}\right].
\end{equation}
\end{subequations}
$e$ is the eccentricity, $a$ is the semi-major axis and $n = 2 \pi/ T $ is the frequency. These quantities are related to the energy and the angular momentum via:
\begin{subequations}
\begin{equation}
\begin{split}
a = \sqrt{\frac{G M}{-2\epsilon}}, \quad e^2 = 1 + 2 l^2 \epsilon , 
\\
\frac{2 \pi}{T} = \frac{\left(-2\epsilon\right)^{3/2}}{G M} = \sqrt{\frac{G M}{a^3}},    
\end{split}
\end{equation}
\end{subequations}
which corresponds to the Keplerian $3^{rd}$ law. Eq.~(\ref{eq:meanAnomNew}) defines the mean anomaly. Two running angles are used to describe the instantaneous position on the ellipse, namely $\theta$ or the true anomaly and $\nu$ or the eccentric anomaly. Fig.~\ref{fig:trueAnomMeanAnom} shows the two anomalies and the relation between these.

\section{Yukawa potential solution}
\label{sec:anaSol}

\begin{figure}[t!]
 	\centering
\includegraphics[width=0.42\textwidth]{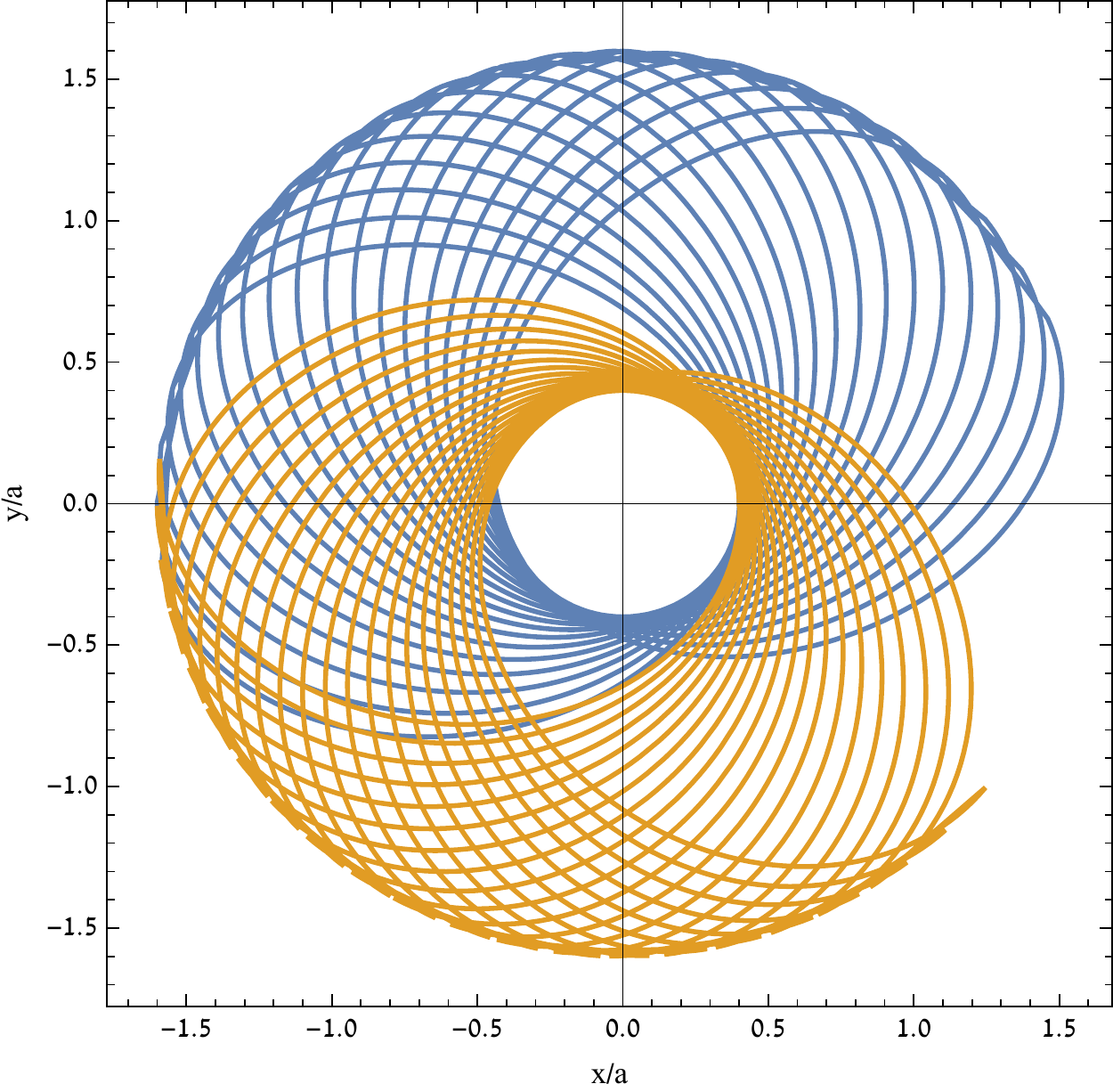}
\caption{\it{Illustration of the analytical solution of the system ~(\ref{eq:timeEq}) and ~(\ref{eq:thetaEqNew}) with different values of positive $\alpha$ (orange) and negative $\alpha$ (blue). The Yukawa potential produces precession also for the extended Newtonian case}}
 	\label{fig:conPlotsNew}
\end{figure}

In this section the solution for the Newtonian potential is generalized via the Yukawa potential. The total energy reduced energy is:
\begin{equation}
\epsilon = \frac{1}{2} \dot{r}^2 - \frac{G M}{r}\left(1 + \alpha \, e^{-m \cdot r}\right) + \frac{l^2}{2 r^2}.
\label{eq:enerYuk}
\end{equation}
In order to find the relation between the energy and the angular momentum to the eccentricity and the semi-major axis, we use edge condition, where the time derivative of the separation is zero:
\begin{equation}
\dot{r}\left[a\left(1 \pm e\right)\right] = 0.
\end{equation}
The separations $a\left(1 \pm e\right)$ are the extremal separations. Inserting this condition into Eq.~(\ref{eq:enerYuk}) yields the relations: 
\begin{subequations}
\begin{equation}
\epsilon = -\frac{G M}{2a}\left[ 1 + \alpha - 2 \alpha  (m \cdot a)  \right],
\end{equation}
\begin{equation}
\frac{l^2}{G M a\left(1 - e^2\right)} = 1 - \alpha -   \frac{f}{2} \left(1 - e^2\right),
\end{equation}
\end{subequations}
where $f = \alpha \, m^2 a^2$. Invoking the condition~(\ref{eq:conSep}) into ~(\ref{eq:enerYuk}) gives a differential equation that relates $\dot{\eta}$ into $\eta$:
\begin{equation}
\frac{1}{\dot{\eta}} = -\frac{1- e \cos\eta}{4} \left(2 (\alpha -2)+\left(e^2-3\right) f+2 e f \cos \eta\right).
\end{equation}
Here, we assume $m\cdot a \ll 1$ and we take the second order correction. The integration gives the elegant solution:
\begin{equation}
\begin{split}
2 \pi \frac{t-t_0}{T} = \eta - e_t \sin \eta + e_r^2 \frac{f}{8} \sin 2 \eta,
\label{eq:timeEq}
\end{split}
\end{equation}
with
\begin{subequations}
\begin{equation}
e_t = \left(1+\frac{2-e_r^2}{4} f\right)e_r,
\end{equation}
\begin{equation}
\begin{split}
T = 2 \pi \sqrt{\frac{a^3}{G M}}  \left[1 -  \frac{\alpha}{2} + \frac{3}{4} f \right] ,   
\end{split}
\end{equation}
\end{subequations}
Using the chain rule over the true anomaly:
\begin{equation}
\begin{split}
\frac{d\theta}{d\eta} = \frac{\dot{\theta}}{\dot{\eta}} = \frac{l/r^2}{\dot{\eta}} = \frac{\sqrt{1-e^2}}{1-e \cos (\eta )} \\\times \left(1 + \frac{\alpha }{2}+f \left(\frac{e^2}{4}-\frac{1}{2} e \cos (\eta )+\frac{1}{4}\right)\right).  
\end{split}
\end{equation}
Solving the integral yields the solution for the true anomaly vs. the mean anomaly:
\begin{equation}
2 \pi \frac{\theta - \theta_0}{\Phi} = \nu + \frac{f}{2} \sqrt{1-e^2} (\eta -\nu ),
\label{eq:thetaEqNew}
\end{equation}
with:
\begin{equation}
\frac{\Phi_{Yuk}}{2 \pi} = 1 + \frac{\alpha}{2}-\frac{f}{4}   \left(1 - e^2-2 \sqrt{1-e^2}\right) .
\label{eq:precYuk}
\end{equation}
Therefore, the the precession rate reads:
\begin{equation}
\Delta\theta_{Yuk} = \frac{\alpha}{2}-\frac{f}{4}   \left(1 - e^2-2 \sqrt{1-e^2}\right) .
\label{eq:precYuk}
\end{equation}
The time eccentricity is modified due to the Yukawa potential presence as in the Post Newtonian correction to the Newtonian potential. The period and the precession are effected as well and could be constraint from the standard solar system planets measurements. Fig.~\ref{fig:conPlotsNew} illustrates the motion of the with positive and negative values of $\alpha$. The sing changes the direction of the motion.

\section{Relativistic Corrections}
\label{sec:relCor}
We examine the motion of a test mass of the Schwarzschild
spacetime. The particle is moving freely on a timelike geodesic of the spacetime. It can be shown that the the relativistic correction changes the angular momentum part in the effective potential by \cite{brouwer1961methods}:
\begin{equation}
\frac{l^2}{2 r^2} \left( 1 -\frac{G M}{r\, c^2} \right).
\end{equation}
The dimensionless ratio $G M/ r\cdot c^2$ is the Schwarzschild correction in the effective potential. Following the same procedure for the modified potential, and taking separations $a\left(1 \pm e\right)$ as an extremal one, yields the relations:  
\begin{subequations}
\begin{equation}
\epsilon = - \frac{G M }{2 a}\left(1 + \alpha -2 \alpha \,    m\cdot a  - \frac{ \beta }{2} (1 + \alpha) \left(1 - e^2\right)\right),
\end{equation}
\begin{equation}
\begin{split}
\frac{l^2}{G M a\left(1 - e^2\right)} =\\ \left(1 + \frac{\beta }{2}  \left(e^2+3\right)\right) \left(1 + \alpha -\frac{f}{2} \left(1-e^2\right)\right),
\end{split}
\end{equation}
\end{subequations}
where we define the dimensionless parameter:
\begin{equation}
\beta = \frac{G M}{a\,c^2 (1-e^2) }.
\end{equation}
For $\beta \ll 1$ the correction agrees with the Post Newtonian correction for massless and spinless body. Invoking the condition~(\ref{eq:conSep}) into the modified energy equation gives a differential equation that relates $\dot{\eta}$ into $\eta$:
\begin{widetext}
\begin{equation}
\begin{split}
\frac{\dot{\eta}^{-1} - \dot{\eta}^{-1}_{(\beta = 0)}}{\sqrt{a^3/G M}} =\frac{\beta }{16}  \left(1 - e^2\right) (e \rho -3) \left(2 (\alpha -2)+f \left(e^2+6 e \rho -7\right)\right).
\end{split}
\end{equation}   
As in the classical case, using the chain rule over the true anomaly:
\begin{equation}
\begin{split}
\frac{d\theta}{d\eta} - \frac{d\theta}{d\eta}_{\left(\beta = 0\right)} = -\frac{\beta  \sqrt{1-e^2}}{16 (e \rho -1)^2} \\\times \left(e \left(e \left(2 (\alpha +2)+\left(e^2+4\right) f\right)+2 e \left(e^2-9\right) f \rho ^2+\rho  \left(14 (\alpha +2)+e^4 f+2 e^2 (\alpha -5 f+2)+41 f\right)\right)-3 (6 (\alpha +2)+7 f)\right). 
\label{eq:thetaetaRel}
\end{split}
\end{equation}
\end{widetext}
The solution for the differential equation yields the same Eq~(\ref{eq:timeEq}) with modified time eccentricity and the period:
\begin{subequations}
\begin{equation}
\frac{e_t}{e} = 1+\frac{2-e^2}{4} f -\frac{ \beta }{2} \left(1 - e^2\right) \left(1 - \frac{2 + e^2}{8} f\right),
\end{equation}
\begin{equation}
\begin{split}
\frac{T}{2 \pi \sqrt{a^3/G M} } = 1 -\frac{\alpha }{2}+\frac{3}{4}f \\+ \frac{3}{16} \beta  \left(1-e^2\right) (4-2 \alpha +7 f).   
\end{split}
\end{equation}
\end{subequations}
The modification for the Yukawa potential modified the time eccentricity only in the second order. Also the leading PN order of the time eccentricity is modified only from the second order. However, the period is modified only also by the first order $\alpha$ both in the Newtonian and the PN expansion terms. 

\begin{figure}[t!]
\centering

\includegraphics[width=0.4\textwidth]{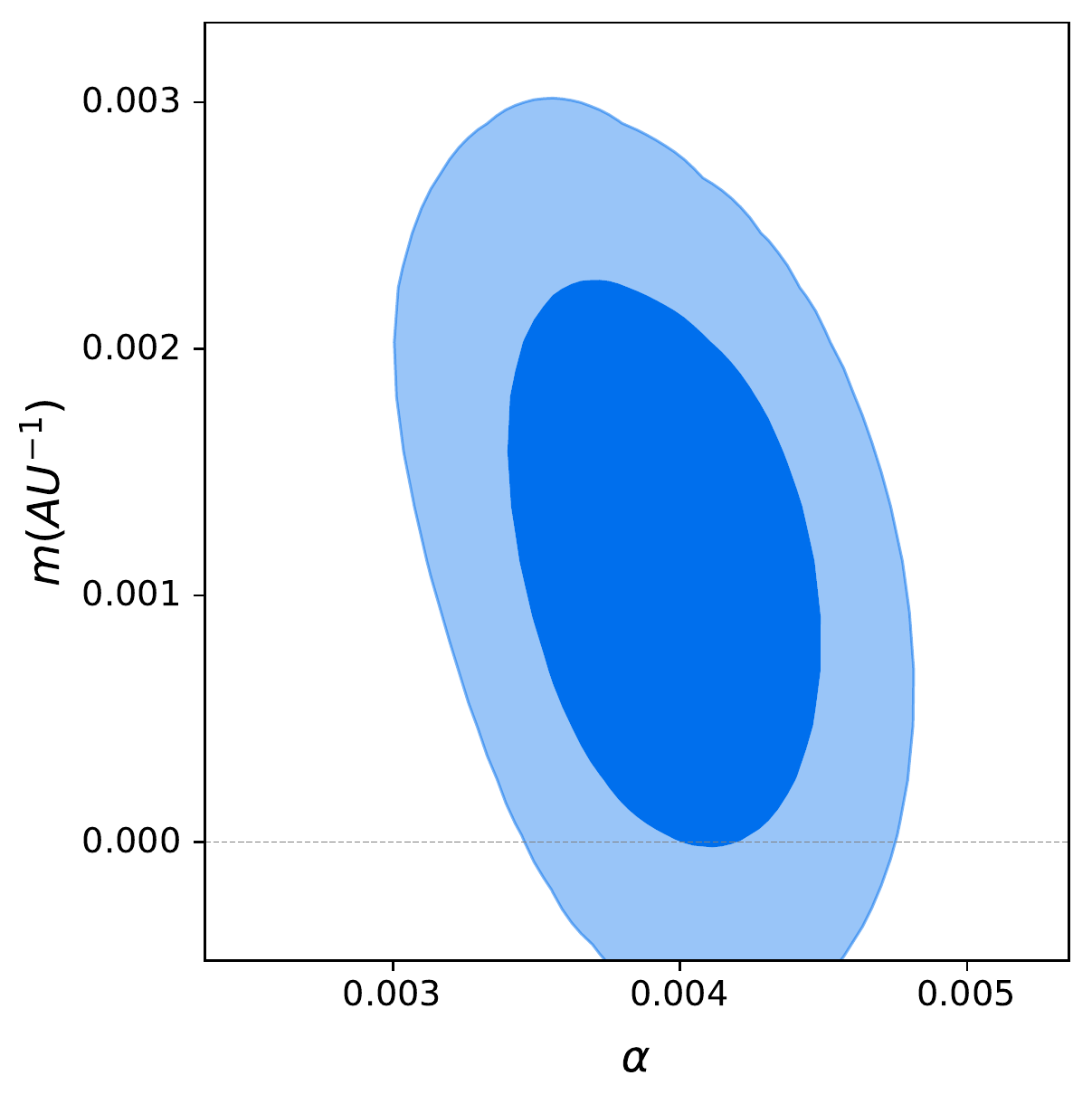}

\begin{tabular}{| c | c | c |}
\hline\hline
Parameter & Solar System & + Cassini \\
\hline\hline  
$\alpha$ &  $\left(3.863 \pm 0.373 \right)\cdot 10^{-3}$ & $\left(4.351 \pm 0.2713 \right)\cdot 10^{-7}$ \\
\hline
$m \, (AU^{-1})$ & $\left(1.066 \pm 0.6074 \right)\cdot 10^{-3}$  & $\left(3.323 \pm 1.627\right)\cdot 10^{-1}$\\
\hline\hline  
\end{tabular} 

\caption{\it{ The contour plot for the Yukawa potential from the solar system precession. The point $(0,0)$ corresponds to GR with no additional interaction.}}
\label{fig:final}

\end{figure}

Solving~(\ref{eq:thetaetaRel}) yields the solution for the true anomaly vs. the mean anomaly:
\begin{equation}
\begin{split}
\frac{2\pi}{\Phi}\left(\theta - \theta_0\right)= \nu + \frac{f}{2} \sqrt{1-e^2} (\eta -\nu )\\ + \frac{\beta}{2} \frac{ e \sqrt{1-e^2} \sin (\eta )}{2-2 e \cos (\eta )}    
\end{split}
\end{equation}
with the preccession:
\begin{equation}
\begin{split}
\Delta \theta_{Yuk, 1 PN} = \frac{\Phi}{2 \pi} - 1 = \Delta\theta_{Yuk} + \Delta\theta_{Yuk, Mod}.    
\end{split}
\end{equation}
where $\Delta\theta_{Yuk, Mod}$ is the modification for the preccession due to the $1^{st}$ order correction. 
\begin{equation}
\begin{split}
\frac{\Delta\theta_{Yuk, Mod}}{\beta} = \frac{e^2+9}{4} 
+\alpha \frac{9+e^2}{8} \\ +\frac{f}{16} \left(3 + e^4-2 \left(\sqrt{1-e^2}+2\right) e^2+18 \sqrt{1-e^2}\right).
\end{split}
\end{equation}
The preccession rate depends also on the dimensionless parameter $\beta$ and also on the Yukawa parameters $\alpha $ and $f$. It is possible to take the first order corrections for the preccession rate and state:
\begin{equation}
\Delta\theta \approx \Delta\theta_{GR} + \Delta\theta_{Yuk}.
\end{equation}
The corrections to the relativistic case also include $\alpha$ and $f$ in the next order correction.

\section{Data Comparison}
\label{sec:dataCom}
In order to complete our discussion on the interactions in two body motion we add the solar system constraint.  \cite{Ip:2015qsa}. The comparison is done with the $\chi^2$ function:
\begin{equation}
\chi^2 = \left(\frac{\Delta\theta_{ob} - \Delta\theta\left(\alpha,m\right)}{\sigma_{\theta}}\right)^2,
\end{equation}
where $\Delta\theta_{ob} \pm \sigma_{\theta}$ is the observed precession of the planets and $\Delta\theta\left(\alpha,m\right)$ is the prediction from the model. We include the precession data of the solar system from \footnote{https://nssdc.gsfc.nasa.gov/planetary/factsheet/}. The combined constraint is approached by using the combined likelihood from the precession measurements from the solar system. 

Regarding the problem of likelihood maximization, we use an affine-invariant Markov Chain Monte Carlo sampler \cite{ForemanMackey:2012ig}, as it is implemented within the open-source packaged $Polychord$ \cite{Handley:2015fda} with the \textbf{GetDist} package \cite{Lewis:2019xzd} to present the results. The prior we choose is with a uniform distribution, where $\alpha \in [0.;1.]$, $m(AU^{-1})\in[0.;1]$.

Fig.~(\ref{fig:final}) shows the preccesion constraint from the solar system constraint. The table below summarizes the final values. We see that adding the solar system constraints, gives very little changer to the modified gravity parameters. However, adding the Cassini bound ($\beta^2 <10^{-5}$) (see~ \cite{Bertotti:2003rm}) reduces the final value of $\alpha$ into $\sim 10^{-7}$ and changes the possible bound value of $m$.

\section{Discussion}
\label{sec:dis}
The Yukawa-like correction to the Newtonian potential is an established result of many modified gravity. This article derives Keplerian-type parametrization for the solution of Yukawa type potential accurate equations of motion for two non-spinning compact objects moving in an eccentric orbit. The modifications are encoded in two parameters: the
strength $\alpha$ and the scale mass $m$ of the Yukawa-term.

In this {\it letter} we used the mean anomaly paramerization with the Yukawa potential and we derive an exact analytical solution  for two body motion. Moreover, we derive an analytical term for the modified Keplerian $3^{rd}$ law and the preccesion terms. With the latest Nasa and Cassini bounds we derive the bounds for the Yukawa strength and mass.

\acknowledgments
D.B gratefully acknowledge the supports of the Blavatnik and the Rothschild fellowships. D.B. acknowledges a Postdoctoral Research Associateship at the Queens' College, University
of Cambridge. We have received partial support from European COST actions CA15117 and CA18108 and STFC consolidated grants ST/P0006811 and ST/T0006941.

\bibliographystyle{apsrev4-1}
\bibliography{ref}

\end{document}